\documentclass[journal]{IEEEtran}
\usepackage[utf8]{inputenc}
\usepackage{gensymb}

\usepackage{ifpdf}

\usepackage{cite}

\ifCLASSINFOpdf
  \usepackage[pdftex]{graphicx}
\else
  \usepackage[dvips]{graphicx}
\fi
\usepackage{amsmath}
\usepackage{bm}
\usepackage{mathtools}
\usepackage{url}

\begin{document}
%
\title{Stochastic Super-Resolution for Downscaling Time-Evolving Atmospheric Fields with a Generative Adversarial Network}
%
%
%
\author{Jussi~Leinonen, Daniele~Nerini and Alexis~Berne
\thanks{J. Leinonen and A. Berne are with the Environmental Remote Sensing Laboratory, École polytechnique fédérale de Lausanne, Lausanne, Switzerland. e-mail: jussi.leinonen@epfl.ch}
\thanks{D. Nerini is with the Federal Office of Meteorology and Climatology MeteoSwiss, Locarno-Monti, Switzerland.}
\thanks{This paper has a video and additional figures available online as supplemental information. Furthermore, the code and datasets that can be used to replicate the results can be found at https://github.com/jleinonen/downscaling-rnn-gan.
}
}

%
%

\markboth{IEEE Transactions in Geoscience and Remote Sensing, 2020}%
{Shell \MakeLowercase{\textit{et al.}}: GAN for Stochastic super-resolution}
%



\maketitle

\begin{abstract}
Generative adversarial networks (GANs) have been recently adopted for super-resolution, an application closely related to what is referred to as ``downscaling'' in the atmospheric sciences: improving the spatial resolution of low-resolution images. The ability of conditional GANs to generate an ensemble of solutions for a given input lends itself naturally to stochastic downscaling, but the stochastic nature of GANs is not usually considered in super-resolution applications. Here, we introduce a recurrent, stochastic super-resolution GAN that can generate ensembles of time-evolving high-resolution atmospheric fields for an input consisting of a low-resolution sequence of images of the same field. We test the GAN using two datasets, one consisting of radar-measured precipitation from Switzerland, the other of cloud optical thickness derived from the Geostationary Earth Observing Satellite 16 (GOES-16). We find that the GAN can generate realistic, temporally consistent super-resolution sequences for both datasets. The statistical properties of the generated ensemble are analyzed using rank statistics, a method adapted from ensemble weather forecasting; these analyses indicate that the GAN produces close to the correct amount of variability in its outputs. As the GAN generator is fully convolutional, it can be applied after training to input images larger than the images used to train it. It is also able to generate time series much longer than the training sequences, as demonstrated by applying the generator to a three-month dataset of the precipitation radar data. The source code to our GAN is available at \url{https://github.com/jleinonen/downscaling-rnn-gan}.
\end{abstract}


%
\IEEEpeerreviewmaketitle

\section{Introduction}
%
%
%
%

\IEEEPARstart{S}{uper-resolution} refers to enhancing the spatial resolution of an image beyond the original resolution. In digital image processing, the term describes various algorithms that take one or more low-resolution images and generate an estimate of a higher-resolution image of the same target \cite{Milanfar2011SR}. In climate science, \emph{downscaling}\footnote{The terminology here is potentially confusing. Upsampling, meaning an operation that increases the number of pixels in an image and thus reduces the physical size of each pixel, is sometimes referred to as ``upscaling'' in image processing. In climate science, the term ``downscaling'' is used instead for an operation that reduces the physical size of pixels and thus improves the resolution. We attempt to avoid this unfortunate contradiction by using the terms ``upsampling'' and ``downsampling'' as they are defined in image processing, and the term ``downscaling'' as it is used in climate science.} is a concept closely related to super-resolution (e.g. \cite{Maraun2018Downscaling,Wood2004HydrologicDownscaling,Fowler2007Downscaling}). It is used especially in connection with precipitation, which can vary sharply over spatial scales of $1\ \mathrm{km}$ or less while global climate models typically have resolutions of tens or hundreds of kilometers. Downscaling bridges this gap by producing precipitation fields at finer resolution for the purpose of assessing the impacts of phenomena such as extreme rainfall.
\IEEEpubidadjcol

Like many other image processing applications, super-resolution has benefited from the introduction of the techniques of deep learning and particularly convolutional neural networks (CNNs). Early attempts at super-resolution using deep CNNs focused on finding image quality metrics that could serve as loss functions that produce sharp images \cite{Dong2016CNNSR,Kim2016DeepSR,Johnson2016PerceptualSR}. More recently, generative adversarial networks (GANs) have been used to train super-resolution CNNs \cite{Ledig2017GANSR,Wang2018ESRGAN}. GANs are a general technique for generating artificial samples \cite{Goodfellow2014GAN} from the training distribution. When used to train CNNs, they can create visually realistic artificial images of, e.g., human faces \cite{Karras2019StyleGAN} and landscapes \cite{Park2019GauGAN}. In super-resolution applications, GANs create reconstructed high-resolution images by using one neural network (the discriminator) to evaluate the quality of the high-resolution outputs, while another network (the generator) is trained to output images that the discriminator considers to be of high quality. The two networks are trained simultaneously against each other (hence ``adversarial'') and thus the discriminator adaptively learns an appropriate reconstruction metric for the dataset rather than relying on expert-provided metrics. The GAN generator may also have a noise input, which the generator learns to map to the variability of the output.

Producing a super-resolution image from only one source image (referred to as single-image super-resolution) is an underdetermined problem that generally does not have a unique solution. Super-resolution techniques therefore try to produce an image that is consistent with the input and that also takes advantage of prior knowledge about the structure of the high-resolution images. Despite the inherent uncertainty in the super-resolution reconstruction, often these methods produce a single output for a given input and rarely estimate the uncertainty of the output. For instance, the state-of-the-art Enhanced Super-Resolution GAN (ESRGAN) architecture does not include a noise input at all and is therefore completely deterministic for a given input  \cite{Wang2018ESRGAN}. This is often acceptable in applications such as enhancing the resolution of natural photographs, where a single plausible solution tends to be sufficient.

In contrast to photograph processing, in climate and weather applications it is crucial to understand and quantify the uncertainty of predictions. Classical precipitation downscaling algorithms have used techniques such as randomized autoregressive models \cite{Rebora2006RainFARM,Terzago2018RainFARM} or multifractal cascades \cite{Lovejoy2006Multifractals} to produce different random realizations of the high-resolution field for a given low-resolution input. GANs offer a natural way to model uncertainty using modern machine-learning methods, less dependent on particular statistical assumptions than the traditional methods. Regardless, the uncertainty aspect has been largely ignored in earlier attempts at improving the resolution of climate fields using deep learning even when employing GANs for this problem (e.g. \cite{Chen2019RadarGANSR}) or for other super-resolution applications related to climate or remote sensing  \cite{Ma2018GANSRRemoteSensing,Jiang2019EdgeEnhancedGAN,Stengel2020AdversarialClimate}, although a few studies have used GANs to represent uncertainty in other atmospheric data problems \cite{Leinonen2019CSMODISGAN,Gagne2020GANParameterization}. Moreover, while GANs have been recently also used to model the time evolution of atmospheric fields \cite{Scher2020RainDisaggGAN}, few studies using deep learning have investigated modeling the uncertainty of the generated high-resolution image in a manner consistent with the time evolution of atmospheric fields --- a problem analogous to video super-resolution, which has also been studied using GANs \cite{Katsaggelos2019VideoSR,Wang2019VideoSR}.

In this paper, we introduce a stochastic super-resolution GAN that can produce an ensemble of plausible high-resolution outputs for a given input. The GAN architecture also includes a recurrent neural network (RNN) structure, which permits the generated outputs to evolve in time in a consistent manner. The architecture is fully convolutional and thus the networks can be trained with small images and later applied to larger ones. We use this GAN to stochastically downscale time series of images from two atmospheric remote-sensing datasets: precipitation measured by the MeteoSwiss ground-based weather radar network, and cloud optical depth imaged by the Geostationary Operational Environmental Satellite 16 (GOES-16). The same architecture is used for both datasets, and thus we expect that the method can be generalized to other atmospheric variables and further applications beyond the atmospheric field.

The rest of this paper is structured as follows. Section~\ref{sect:methods} describes the network architecture and training as well as the validation of the results. Section~\ref{sect:data} describes the datasets and their preprocessing, and Section~\ref{sect:results} presents and discusses the evaluation results. Finally, Section~\ref{sect:conclusions} concludes the paper and presents objectives for future work.

\section{Methods} \label{sect:methods}

\subsection{Overview}

A GAN consists of two neural networks: the generator ($G$) and the discriminator ($D$). The discriminator is trained to determine whether or not its input is an example from the training dataset, while the generator is simultaneously trained to produce artificial samples that the discriminator classifies as real. Thus, the generator learns to produce realistic-looking artificial samples. In this study, we use a \emph{conditional} GAN \cite{Mirza2014ConditionalGAN}, in which both $G$ and $D$ are given an additional condition. In the case of super-resolution, the condition is a low-resolution image, and the discriminator is trained to distinguish between real high-resolution images from the training dataset and artificial high-resolution images produced by the generator, conditionally to the corresponding low-resolution images. 

For additional background on GANs, we refer the reader to \cite{Foster2019GenerativeDeepLearning}, while a general overview of deep-learning methods can be found in \cite{Goodfellow2016DeepLearning}.

\subsection{Network architecture} \label{sect:architecture}

In our GAN, both $G$ and $D$ are deep CNNs which make extensive use of residual blocks \cite{He2016ResNet}. The residual blocks process their input through two activation and convolution layers and finally add the input to the output at the end of processing. Consequently, an inactive residual block (one with near-zero weights in the convolutional layers) acts as an identity map. Thus, the number of residual blocks in a network is often flexible since the blocks that the network does not use simply pass their input through. As training progresses, residual networks may activate additional blocks as the network learns to take advantage of deeper features. The numbers of residual blocks in our networks were determined by an iterative design process but, for the above-mentioned reasons, their exact number is not critically important as having too many residual blocks need not be harmful to performance, although it does increase computational cost.

In contrast to most GANs, our networks also employ recurrent layers in the form of convolutional gated recurrent units (ConvGRUs), variants of the gated recurrent unit (GRU) \cite{Cho2014GRU}. ConvGRUs replace learned affine transforms in the standard GRU with two-dimensional convolutions. ConvGRU layers learn the appropriate update rules from one time step to the next, enabling the GAN generator to model the evolution of the fields with time, and allowing the discriminator to evaluate the plausibility of image sequences rather than single images. These layers, along with the closely-related convolutional long short-term memory (LSTM) layers, have been previously applied to modeling the time evolution of precipitation fields \cite{Shi2015ConvLSTMPrecip,Tian2020GRUNowcasting}.

\begin{figure*}[!t]
\centering
\includegraphics[width=0.8\textwidth]{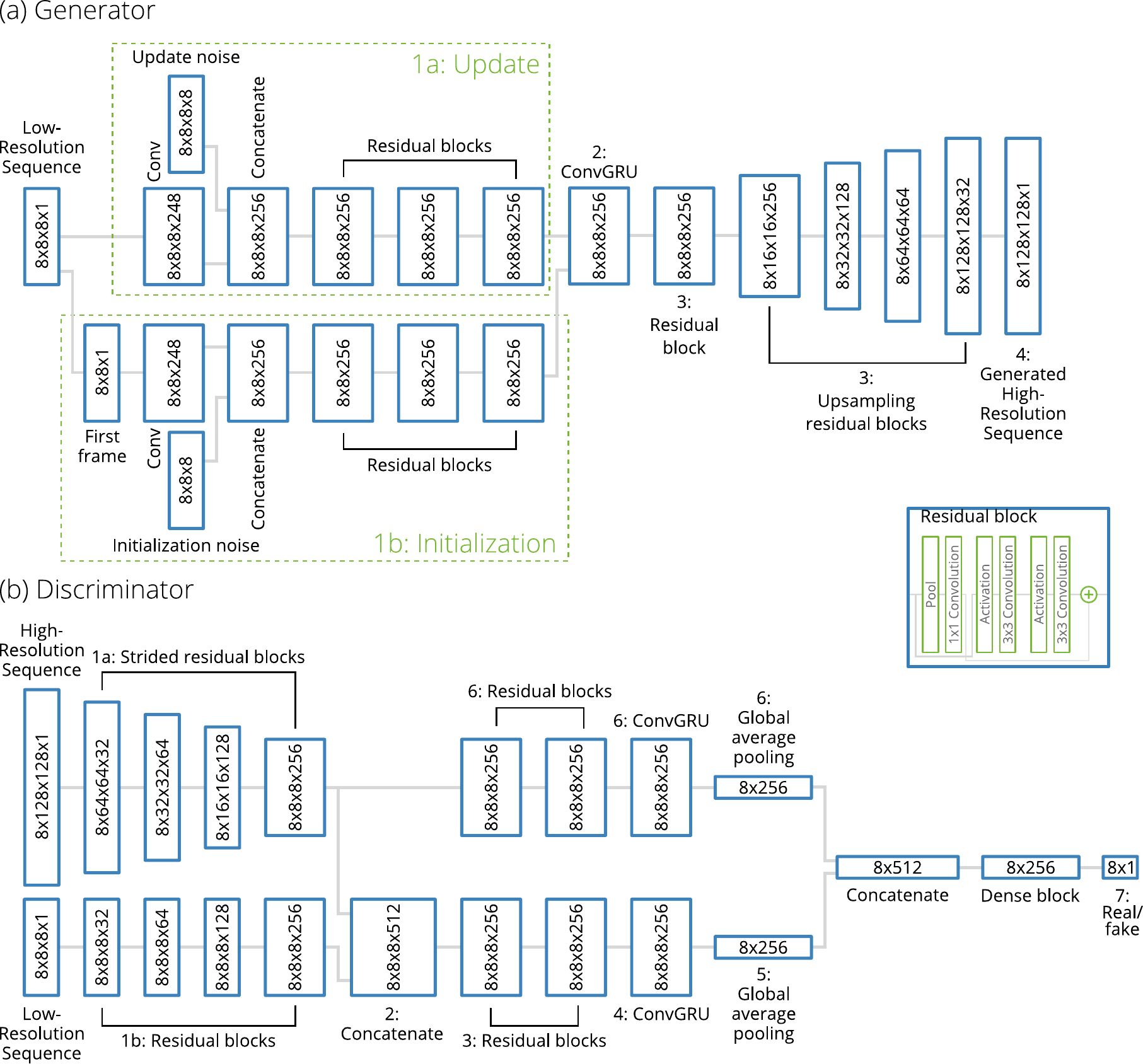}
\caption{The architectures of (a) the generator and (b) the discriminator. The numbered labels correspond to the descriptions in Section~\ref{sect:architecture}. The dimensions shown here are for the training configuration where high-resolution image size $h \times w = 128 \times 128$, the number of frames per sequence $N_\mathrm{t}=8$ and the number of variables $N_\mathrm{v}=1$ for both datasets considered here. After training, the network can be evaluated using different values of these parameters.}
\label{fig:networks}
\end{figure*}

The architectures of our $G$ and $D$ networks are shown in Fig.~\ref{fig:networks}. Below, we give brief descriptions of the organization of the networks; the exact implementation using TensorFlow  \cite{Abadi2016TensorFlow} and tf.keras \cite{Chollet2015Keras}, which is TensorFlow's high-level API for building and training deep learning models, can be found in the source code published at \url{https://github.com/jleinonen/downscaling-rnn-gan}.

The generator $G$ starts with a time series of low-resolution fields (the conditioning variable), given as a 4-D tensor of dimension $N_\mathrm{t} \times h \times w \times N_\mathrm{v}$, where $N_\mathrm{t}$ is the number of time steps, $h$ and $w$ are the pixel height and width of the image, respectively, and $N_\mathrm{v}$ is the number of variables. The time steps are assumed to be at constant intervals, and the size of one pixel is assumed to always correspond to a constant, well-defined physical size. The time series is processed through the following steps of the network:
\begin{enumerate}
\item \textit{Encoding}: \label{item:gen-encoding} 
  \begin{enumerate}
    \item The low-resolution input tensor is mapped to a larger number of channels using a convolutional layer, and concatenated with the noise input using a different noise instance for each time step. This data is then processed through a series of three residual blocks. The inputs are thus encoded into a deep representation. \label{item:gen-encoding-a}
    \item Using a similar series of layers as with step~\ref{item:gen-encoding-a} above but with independent weights and only for the first time step, the initial state of the recurrent layer is derived. \label{item:gen-encoding-b}
  \end{enumerate}
\item \textit{Recurrence}: The time evolution of the deep representation of the field is modeled with a ConvGRU layer. The input to the ConvGRU layer is the result of step~\ref{item:gen-encoding-a} above, while the initial state is derived from step~\ref{item:gen-encoding-b}.
\item \textit{Decoding/upsampling}: The result of the ConvGRU layer is processed through a series of alternating residual blocks and upsampling layers. Each upsampling operation increases both spatial dimensions by a factor of two, using bilinear interpolation on the hidden representation. The residual blocks process the information to a less deep level of representation. We use four upsampling blocks, resulting in a resolution enhancement by a factor of $K=16$. Different numbers of upsampling blocks could be used to obtain different factors of $K=2^N$ with $N$ a positive integer, but this would require retraining the GAN, requiring increased computation time for training, and hence we concentrate on $K=16$ in this work. \label{item:gen-upsampling}
\item \textit{Output}: The output of the last hidden layer is mapped using one final convolution to a high-resolution tensor of dimension $N_\mathrm{t} \times Kh \times Kw \times N_\mathrm{v}$. A sigmoid activation constrains the final output between $0$ and $1$.
\end{enumerate}
$L_2$ weight regularization is used in the generator. All non-recurrent layers use shared weights for each time step; this allows the generator to operate with any number of time steps. The generator has approximately $13.6$ million trainable weights.

The discriminator $D$ starts with a pair of high- and low-resolution sequences. The task of the discriminator is to determine whether or not these are a pair originating from the training dataset. The processing steps below are used to achieve this:
\begin{enumerate}
\item \textit{Encoding/downsampling}:
\begin{enumerate}
\item The high-resolution input is processed using a series of three residual blocks that use strided convolutions to downsample the input and encode it into a deep representation. As with the generator, the same weights are used for each time step. \label{step:disc-encoding-a}
\item The low-resolution input is processed identically to step~\ref{step:disc-encoding-a}, except the convolutions are not strided and thus no downsampling is performed. As a result, the output has the same dimensions as that of step~\ref{step:disc-encoding-a}. \label{step:disc-encoding-b}
\end{enumerate}
\item \textit{Combination}: The outputs of steps~\ref{step:disc-encoding-a} and~\ref{step:disc-encoding-b} are concatenated. \label{step:disc-combination}
\item \textit{Further encoding}: The joint output from step~\ref{step:disc-combination} is processed through two residual blocks for additional encoding. \label{step:disc-joint-encoding}
\item \textit{Recurrence}: The time consistency of the field is evaluated with a ConvGRU layer; unlike with the generator we simply initialize the state to zeros.
\item \textit{Global average pooling}: The average of each feature map is taken, pooling the activations at the different locations. \label{step:disc-pooling}
\item \textit{High-resolution processing}: We also process the output of step~\ref{step:disc-encoding-a} separately through steps~\ref{step:disc-joint-encoding}--\ref{step:disc-pooling} using independent weights. The motivation for this branch is to evaluate the quality of the high-resolution image separately from the consistency of the low/high-resolution pair. \label{step:disc-hr}
\item \textit{Output}: The results of steps~\ref{step:disc-pooling} and~\ref{step:disc-hr} are concatenated. The result is processed through one more fully connected layer, then mapped to $N_\mathrm{t}$ scalar values.
\end{enumerate}
Spectral normalization \cite{Miyato2018SpectralNormalization} is used to constrain the discriminator. The number of trainable weights in the discriminator is approximately $15.1$ million.

Leaky rectified linear unit (ReLU) activations \cite{Maas2013RectifierNonlinearities} with negative slope $0.2$ are used in both $G$ and $D$ except for the update and initialization networks in $G$ (items \ref{item:gen-encoding-a} and \ref{item:gen-encoding-b} in the description of $G$), which use regular ReLU activations. Using the regular ReLU in these parts of the network proved useful for improving stability when the generator is evaluated over long time series; we speculate that this is because the ReLU activation can become completely inactive while the leaky ReLU cannot. Meanwhile, using leaky ReLU in the upsampling part of $G$ (item \ref{item:gen-upsampling} in the description) produced fewer artifacts than regular ReLUs. 

\subsection{Training}
Formally, the conditional GAN optimization objectives are
\begin{IEEEeqnarray}{C}
\min_{\bm{\theta}_D} \mathrm{E}_{\mathbf{x},\mathbf{y},\mathbf{z}} \left [ L_D(\mathbf{x},\mathbf{y},\mathbf{z};\bm{\theta}_D) \right ] \label{eq:wgan-discriminator-loss} \\
\min_{\bm{\theta}_G} \mathrm{E}_{\mathbf{y},\mathbf{z}} \left [ L_G(\mathbf{y},\mathbf{z};\bm{\theta}_G) \right ] \label{eq:wgan-generator-loss}
\end{IEEEeqnarray}
where $\mathbf{x}$ represents real samples (for us, high-resolution sequences), $\mathbf{y}$ represents the condition (low-resolution sequences) and $\mathbf{z}$ is the noise. We denote the discriminator loss as $L_D$, the generator loss as $L_G$, and the corresponding trainable weights as $\bm{\theta}_D$ and $\bm{\theta}_G$ respectively.
We trained our GAN as a Wasserstein GAN with gradient penalty (WGAN-GP) \cite{Guljarani2017WGANGP}, using a gradient penalty weight of $\gamma=10$. The combined conditional WGAN-GP losses for $D$ and $G$ are
\begin{IEEEeqnarray}{rCl}
L_D(\mathbf{x},\mathbf{y},\mathbf{z};\bm{\theta}_D) &=& D(\mathbf{x},\mathbf{y}) - D(G(\mathbf{y},\mathbf{z}),\mathbf{y}) + \nonumber \\
&& \gamma (||\nabla_{\hat{\mathbf{x}}} D(\hat{\mathbf{x}},\mathbf{y})||_2 - 1)^2 \\
L_G(\mathbf{y},\mathbf{z};\bm{\theta}_G) &=& D(G(\mathbf{y},\mathbf{z})).
\end{IEEEeqnarray}
where the samples $\hat{\mathbf{x}}$, used to compute the gradient penalty term, are randomly weighted averages between real and generated samples:
\begin{equation}
\hat{\mathbf{x}} = \epsilon \mathbf{x} + (1-\epsilon)G(\mathbf{y},\mathbf{z})
\end{equation}
with $\epsilon$ sampled randomly from the uniform distribution between $0$ and $1$. Intuitively, the Wasserstein loss can be understood as the discriminator trying to make its output as large as possible for generated samples and as small as possible for real samples. The gradient penalty acts to constrain the discriminator output, which is otherwise unbounded.

As the optimization goals in Eqs.~\ref{eq:wgan-discriminator-loss} and~\ref{eq:wgan-generator-loss} are contradictory, $D$ and $G$ must be trained adversarially. We alternated between training $D$ with five batches and $G$ with one, a strategy that was generally found to be beneficial by \cite{Kurach2019LargeScaleGAN}. We used a batch size of $16$, determined by the amount of memory available in the graphics processing unit (GPU). The Adam optimizer \cite{Kingma2014Adam} was used for most of the optimization, with a learning rate of $10^{-4}$ for both $G$ and $D$. We found that Adam converged quickly to reasonable image quality but the solutions tend to oscillate, even with reduced learning rates. Therefore, near the end of the training after $350000$ training sequences, we switched to stochastic gradient descent (SGD) with a learning rate of $10^{-5}$.

The generator was trained with $400000$ sequences, corresponding to $3.2$ million individual images, and the discriminator with 2 million sequences (10 million images). This corresponded to roughly $48$ hours for each application using an Nvidia P100 GPU. Sample diversity was increased by using random rotation (by $0\degree$, $90\degree$, $180\degree$ or $270\degree$) and random mirroring on the image time series. This makes the GAN approximately invariant with respect to $90\degree$ rotations in addition to the translation and time invariance that are features of the network design.

\subsection{Validation and tuning} \label{sect:validation}

While GANs are expected to converge towards the underlying data distribution of their input dataset, frequently (e.g. \cite{Karras2018ProgressiveGAN}) they do not reproduce enough variability. There has been progress in quantifying the quality and variability of generated samples for unconditional GANs using metrics like the Frechet Inception Distance (FID) \cite{Heusel2017FID}, but the FID is not directly applicable to the type of conditional GAN considered here because the training dataset generally contains only one output for each input and therefore the underlying distribution cannot be reliably estimated.

As a simple metric of image quality, we use the root-mean-square error \begin{equation}
\mathrm{RMSE} = \sqrt{\frac{1}{N} \sum_{i=1}^N \left (x_{\mathrm{real},i} - x_{\mathrm{gen},i} \right )^2},
\end{equation}
where $x_i$ are the individual pixel values of the real image, $x_{\mathrm{gen},i} = G(\mathbf{y},\mathbf{z})_i$ are the corresponding pixels of the generated image, and $N$ is the number of pixels. To evaluate if the generated images properly reproduce the spatial structure of the true images, we also compute the multi-scale structural similarity index (MS-SSIM), defined in \cite{Wang2003MSSSIM}, and the log spectral distance (LSD) which gives the difference of the power spectra in decibels (dB):
\begin{equation}
\mathrm{LSD} = \sqrt{\frac{1}{N} \sum_{i=1}^N \left ( 10 \log_{10} \frac{P_{\mathrm{real},i}}{P_{\mathrm{gen},i}}  \right )^2 } 
\end{equation}
where $P_\mathrm{real}$ and $P_\mathrm{gen}$ are the power spectra of the real and generated images, respectively.

For assessing whether the GAN generates the correct amount of variability, we propose to adapt a rank-statistics approach from ensemble weather forecasting \cite{Talagrand1997Evaluation,Candille2006PredictionEvaluation} to obtain a heuristic measure of the variability of the sequences produced by the conditional GAN. The underlying concept is as follows. For each sample we have a single ``ground truth'' (the real high-resolution sequence) and an ensemble of $N_\mathrm{p}$ predictions (we can generate as many predictions as we wish by re-evaluating the GAN with different instances of the noise). Then, for each pixel in the image we can define the \emph{normalized rank} of the actual value among all $N_\mathrm{p}$ predictions as $r = N_\mathrm{s}/N_\mathrm{p}$, where $N_\mathrm{s}$ is the number of predictions in the ensemble for which the value of that pixel is smaller than the corresponding ground-truth pixel (the rank is randomized for ties). Clearly $0 \leq r \leq 1$, and if the sample is from the same distribution as the predictions, $r$ should be uniformly distributed over this range when averaged over many pixels and many sequences. Consequently, we can use the uniformity of the distribution of $r$ as an evaluation metric for the correct variability of the generated images.

The distribution of $r$ can be evaluated visually by examining the histogram of $r$, as demonstrated by, e.g., \cite{Hamill2001RankHistograms}. We can also quantify the uniformity with various distribution distance metrics between the rank distribution $P_r$ and the uniform distribution over the possible values of $r$ (since we take a finite sample of predictions, the possible values are discrete). Here, we investigate several such metrics. First, the Kolmogorov--Smirnov (KS) statistic \cite{Dekking2005Probability} between two sets of probabilities $P$ and $Q$ is defined as
\begin{equation}
\mathrm{KS} = \sup \left | C-D \right |
\end{equation}
where $C$ and $D$ are the cumulative distribution functions (CDFs) of $P$ and $Q$, respectively. Second, the Kullback--Leibler divergence ($D_\mathrm{KL}$) \cite{Kullback1951DKL} of $P$ with respect to $Q$ is
\begin{equation}
D_\mathrm{KL}(P||Q) = \sum_i P(r_i) \log \left ( \frac{P(r_i)}{Q(r_i)} \right )
\end{equation} 
where $r_i$ are the different values that the rank can attain. Unlike KS, $D_\mathrm{KL}$ is generally not symmetric between $P$ and $Q$. Typically, $P$ denotes the ``ideal'' distribution and $Q$ an approximation, so in this work we use the uniform distribution for $P$ and the observed rank distribution for $Q$. As the KS statistic measures the distance of the CDFs and $D_\mathrm{KL}$ relates to the information content difference of the probabilities, these two statistics capture different aspects of the differences between the rank distribution and the uniform distribution. We also compute the outlier fraction (OF), also called outlier percentage (OP) when given in percent units, which is defined as the fraction of ground-truth samples that lie outside the ensemble of predictions.

Using the complete ensemble, we can also evaluate the image quality with a metric that utilizes the entire ensemble of predictions, the continuous ranked probability score (CRPS) \cite{Gneiting2007ProperScoring}. For a given pixel, CRPS is defined  as the integral of the squared difference of the CDF of the ensemble members (denoted as $F$) and the CDF of the observations. For a single observation (the pixel $x_{\mathrm{real},i}$ from the real image), the observation CDF is a Heaviside step function $H$ shifted to the point $x_{\mathrm{real},i}$, giving CRPS for the pixel $i$ as
\begin{equation}
\mathrm{CRPS} = \int_{-\infty}^{\infty} \left ( F(x')-H(x'-x_{\mathrm{real},i}) \right )^2 \mathrm{d}x'.
\end{equation}
The CRPS for an entire image is obtained as the mean of the pixelwise CRPS scores. CRPS can be understood as a generalization of the mean absolute error, to which it is reduced if there is only one ensemble member.

In this paper, all of the above-mentioned metrics are calculated for the data transformed to the $[0,1]$ range as explained in Sect.~\ref{sect:data}.

\section{Data} \label{sect:data}

To demonstrate that the network can learn the structures of different types of atmospheric fields, we trained it independently with two datasets. The first was a collection of samples drawn from the MeteoSwiss weather radar composite \cite{Germann2015MCHComposite} over the year 2018 (hereafter referred to as the ``MCH-RZC'' dataset). The samples were selected and processed as described in \cite{Leinonen2019ClimateGAN} and released in \cite{Leinonen2019WeatherRadarLearning}. The dataset contains $180000$ image sequences, each of which consists of $8$ images of $128 \times 128$ pixel size, each pixel corresponding to a physical size of $1\ \mathrm{km}$. The time interval between subsequent images in a sequence is $10\ \mathrm{min}$. The image size and the number of images in each sequence were chosen as a compromise between the amount of training data and the available computational resources.  The pixel values express the precipitation rate $R$ in units of $\mathrm{mm\ h^{-1}}$; this has been derived from the radar reflectivity, quality controlled, and corrected for various biases. We preprocessed the RZC data by taking the logarithm of $R$, which leads to a regular distribution since $R$ is known to have a near-lognormal distribution for $R>0$ \cite{Kedem1987Lognormality}, making learning easier. The $R=0$ case will be discussed later in this section.

The other dataset is derived from the cloud optical thickness $\tau$ observed by the GOES-16 satellite \cite{Heidinger2020GOESRCloud} (we refer to this dataset as ``GOES-COT'' in the rest of the paper). We used data from April--December 2019, the period after GOES-16 full-disk scans were switched to Mode 6 which provides data every $10\ \mathrm{min}$ (which is only coincidentally the same as with the MCH-RZC dataset; any time interval would work). As the cloud optical thickness is only available at daytime and its accuracy can be affected by high solar zenith angles, we limited the data use to hours between 14 UTC and 20 UTC, corresponding to approximately 09 to 15 local solar time at the sub-satellite point. From these data, we randomly extracted $108544$ image time series of the same dimensions as the weather radar data. The geometric distortion caused by the Earth's curvature and the satellite point of view was corrected by projecting the data to orthographic projection \cite{Snyder1987Projections} with a spatial resolution of $2\ \mathrm{km}$ per pixel. In order to minimize distortion, the sampling was constrained to a box bounded by $30\degree\mathrm{S}$ and $30\degree\mathrm{N}$ latitude, and $105\degree\mathrm{W}$ and $45\degree\mathrm{W}$ longitude (the center of the longitude range being the sub-satellite point at $75\degree\mathrm{W}$). As with the precipitation dataset, we took the logarithm of $\tau$ to make the distribution more even, following \cite{Leinonen2016CloudRetrieval}.

The image given to the GAN during training and evaluation is a transformed variant of the variable $x$ (where $x$ can be either $\log(R)$ or $\log(\tau)$). While the distribution of both variables becomes smoother with the logarithmic transformation, it necessitates special processing in empty (non-precipitating or non-cloudy) regions where the logarithm is not defined. We solve this with the following transformation: Empty pixels are mapped to $0$ and the detectable range $[x_\mathrm{min}, x_\mathrm{max}]$ is shifted and scaled to $[\theta,1]$, thus transforming the entire dataset to $[0,1]$. The threshold $\theta$ is a small positive value that separates the non-precipitating values from the precipitating ones. The transformation is reversible, and consequently when postprocessing the GAN-generated fields we consider every pixel with a value below $\theta$ as empty while values larger than $\theta$ are mapped back to $x$. We used $\theta \approx 0.17$ for both datasets, and did not find the results particularly sensitive to the choice of this parameter. To suppress artifacts that would sometimes appear at the sharp edges caused by the thresholding, we smoothen the images with a Gaussian filter before feeding them to the network. This filter also has the effect of inhibiting certain artifacts in the MCH-RZC dataset that occasionally result from processing the data from multiple radars into a single composite on a regular Cartesian grid.

Each low-resolution image is obtained from its high-resolution counterpart by taking the average of the linear (not logarithmic) values of $R$ or $\tau$ for each non-overlapping $16 \times 16$ pixel tile in the image, then applying the logarithmic transformation and the mapping to $[0,1]$ as described above. Due to the averaging process, some of the averaged pixels may initially have values between $0$ and $\theta$; these are truncated to $0$ in order to prevent the GAN from taking advantage of data that are invisible in the visualizations.

To ensure that we avoid the scenario where the GAN simply memorizes the training set, we set aside $10\%$ of samples, randomly selected, from each dataset to be used as the validation set. The samples from the validation set were not used for training but were used to monitor the progress of the training. Furthermore, to examine how well the GAN generalizes to data that is not sampled from exactly the same data as the training set, we constructed test datasets of $1024$ samples for both data sources using data from a different time period. For MCH-RZC, the test data were selected from the year 2017, while for GOES-COT they were sampled from the April 1--20, 2020 time period. Except where mentioned otherwise, all visualizations shown in this paper were generated using the test sets, ensuring that the examples are from data that the GAN has not been exposed to during training.

\section{Results} \label{sect:results}

\subsection{Examples of generated sequences} \label{sect:examples}

\begin{figure}[!t]
\centering
\includegraphics[width=\linewidth]{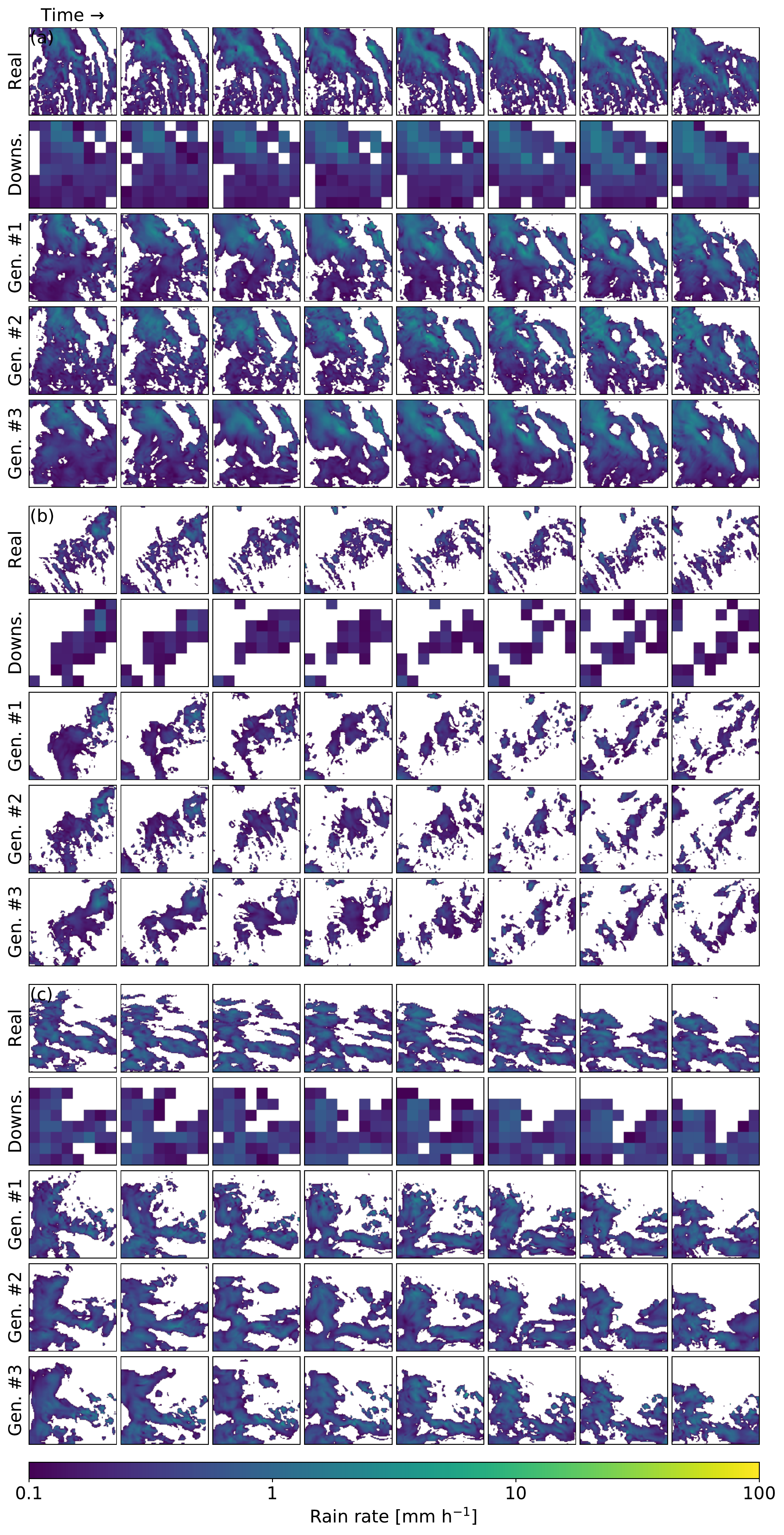}
\caption{Examples of reconstructed image sequences from the MCH-RZC test dataset. Each main panel (a)--(c) shows the real high-resolution image on the top row, the downsampled version on the second row, and three examples of reconstructions created by the GAN on the last three rows.}
\label{fig:examples-mchrzc}
\end{figure}

We show three examples of GAN-reconstructed time series from the MCH-RZC test dataset in Fig.~\ref{fig:examples-mchrzc}. These were generated using the generator saved after $361600$ training sequences, selected based on the metrics shown in Sections~\ref{sect:quality} and~\ref{sect:variability} as well as a subjective check of the quality and stability of the generated sequences. For each example, Fig.~\ref{fig:examples-mchrzc} shows the true high-resolution sequence, the $16 \times 16$ downsampled sequence, and three different reconstructions. The first example (Fig.~\ref{fig:examples-mchrzc}a) shows a region with different rainfall structures in different parts of the image, with a relatively uniform structure at the top center and a highly spatially variable structure at the bottom. At the top, all three reconstructions produce similar, uniform structure that strongly resembles the texture of the original. Meanwhile, we see a significant difference in the structure of the cells developing at the bottom where reconstructions \#1 and \#2 produce much more granular structures than reconstruction \#3, which creates a much more uniform structure at the bottom. This example demonstrates how the difference in granularity remains consistent over time: \#1 and \#2 more spatially variable than \#3 for all time steps. In Fig.~\ref{fig:examples-mchrzc}b, the precipitation is organized over very short scales everywhere in the image. The structure and orientation of the generated cells varies between the reconstructions. None of the three generated examples captures exactly the orientation of the original cells, the information from which is lost in the downsampling process. Regardless, the GAN can clearly infer the type and scale of precipitation cells fairly accurately from the low-resolution image and produce different guesses about the underlying structure. The last example, in Fig.~\ref{fig:examples-mchrzc}c, shows another complex scene that contains different structures in different parts of the image. Here, too, it can be seen that the GAN can generate different solutions for a given scene: The overall structure is the same in all reconstructions, but the details are quite different. 

\begin{figure}[!t]
\centering
\includegraphics[width=\linewidth]{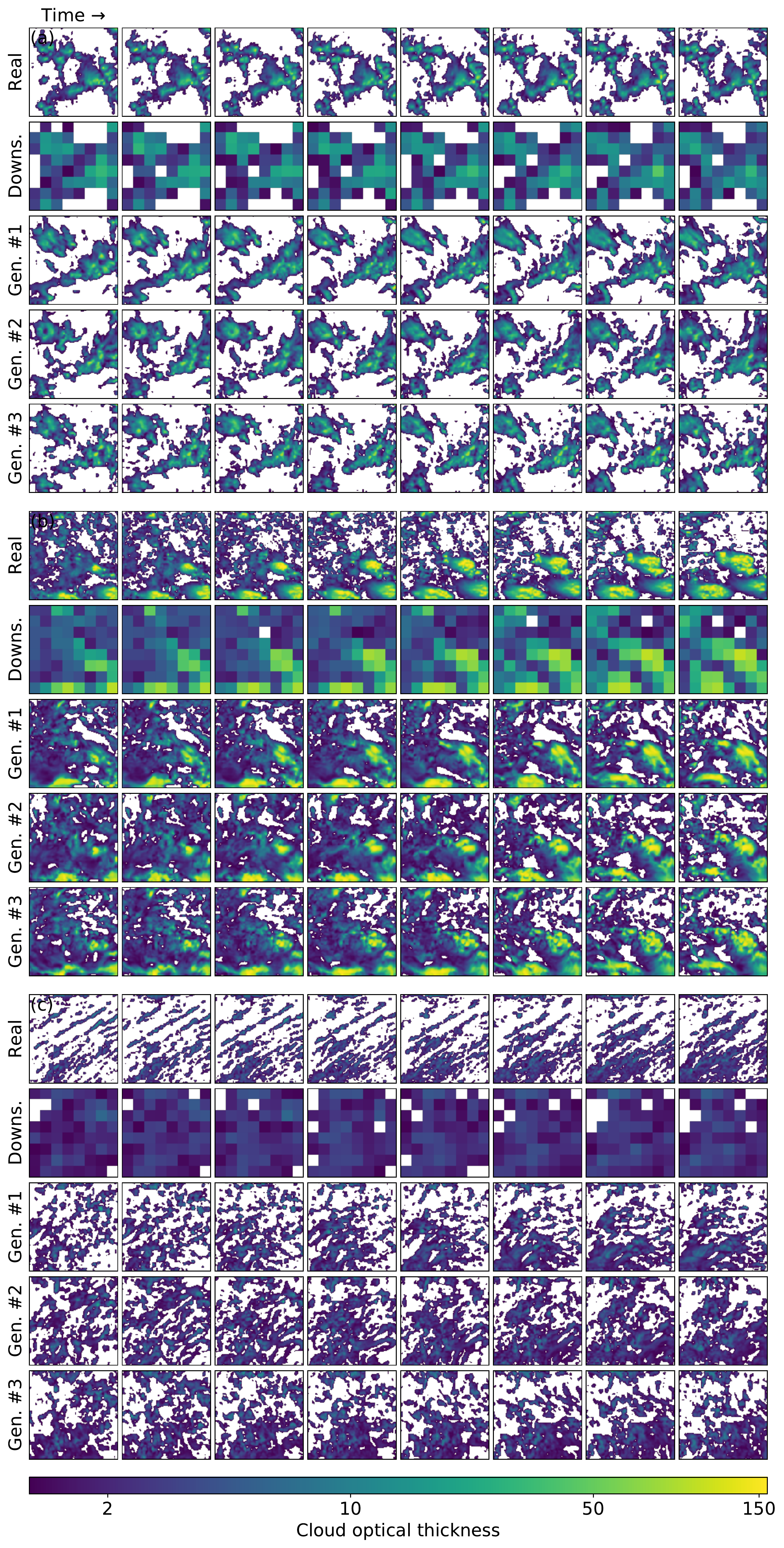}
\caption{As Fig.~\ref{fig:examples-mchrzc}, except for the GOES-COT test dataset.}
\label{fig:examples-goescod}
\end{figure}

Fig.~\ref{fig:examples-goescod} displays three examples for the GOES-COT test dataset, using the generator obtained after $371200$ training sequences. These data generally have more intricate texture than the MCH-RZC dataset, with patterns occurring over shorter scales. This is partially a result of the different spatial resolutions of the datasets, $1\ \mathrm{km}$ for MCH-RZC and $2\ \mathrm{km}$ for GOES-COT. The case of Fig.~\ref{fig:examples-goescod}a has very strong contrasts in the cloud optical thickness, sometimes occurring over distances of only a few pixels. These contrasts are lost in the downsampling; regardless, the GAN is able to generate a pattern at an approximately correct scale and spatial structure. The reconstructions differ in terms of the exact location of the generated clouds, reflecting the uncertainty of the GAN about the correct solution.  Fig.~\ref{fig:examples-goescod}b shows another case of highly complex cloud organization with high COT maxima and strong contrasts over short distances. This example demonstrates the time consistency of the solutions particularly well; for example, the empty regions are in different locations in the different images but their location remains consistent from one time step to the next. Furthermore, there are differences in the texture of the clouds between the generated images: for example, the high-COT region in the center right of the last few frames contains a cell structure in reconstruction \#3, while it is more uniform in reconstructions \#1 and \#2.  Finally. Fig.~\ref{fig:examples-goescod}c shows a highly anisotropic case where the clouds have a strongly preferred orientation in the original high-resolution image. The GAN has some difficulty inferring the correct orientation, which is lost in the downsampling, and generates fairly different solutions to reflect its uncertainty of the correct answer. Some solutions in the corresponding figure in the supplement (\texttt{examples-goescod-random-02.pdf}) include even more strongly oriented clouds, although none match the correct solution exactly. The generated clouds in reconstruction \#2 exhibit some preferred orientation.

We selected the examples in Figs.~\ref{fig:examples-mchrzc}--\ref{fig:examples-goescod} manually in order to illustrate the behavior of the network in different cases. As such, they are a limited and non-representative sample of the datasets. Moreover, it is impossible to convey the full variability of the generated solutions using only the three ensemble members that we are limited to because of space constraints. To address this issue, we have included more examples, randomly selected from the test datasets and with more ensemble members generated with the GAN, in the supplement available online alongside this article.

\subsection{Reconstruction quality} \label{sect:quality}

\begin{figure}[!t]
\centering
\includegraphics[width=\linewidth]{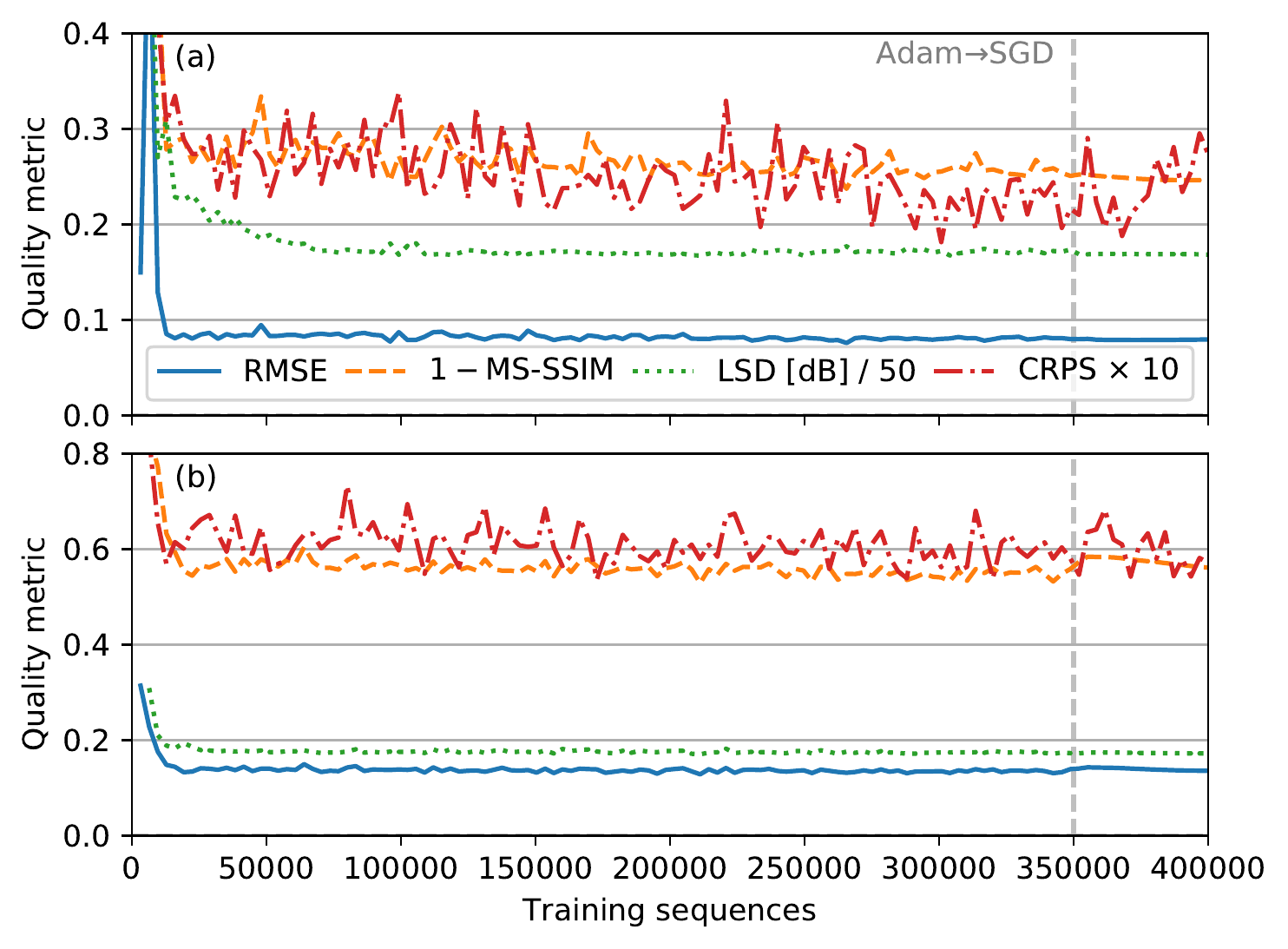}
\caption{Metrics of the image quality from the GAN-generated ensemble. The blue solid line shows the RMSE, the orange dashed line shows $1-$MS-SSIM, the green dotted line shows the LSD (divided by $50$ to bring it to a similar scale as the other metrics), and the red dash-dotted line shows the CRPS multiplied by $10$. Panel (a) shows the results for the MCH-RZC validation dataset and panel (b) for the GOES-COT validation dataset.}
\label{fig:quality-metrics-time}
\end{figure}

To assess the development of image quality as the GAN is trained, we computed the RMSE, MS-SSIM, LSD and CRPS metrics, as described in Section~\ref{sect:validation}, at intervals of $3200$ generator training sequences. All of these metrics were calculated for the data transformed to the $[0,1]$ range as explained in Sect.~\ref{sect:data}. The evolution of the average of these metrics over a sample drawn from the validation dataset is shown in Fig.~\ref{fig:quality-metrics-time}. The numbers for the fully trained GAN are shown in Table~\ref{table:metrics} for both the test and validation datasets.
\begin{table*}[!t]
    \centering
    \begin{tabular}{|l|c|c|c|c|c|c|c|c|}
        \hline
        &  RMSE & MS-SSIM & LSD (dB) & CRPS & KS & $D_\mathrm{KL}$ & OF & Mean rank\\
        \hline
        GAN, MCH-RZC valid. & $0.079$ & $0.750$ & $8.445$ & $0.020$ & $0.029$ & $0.014$ & $0.046$ & $0.502$ \\
        GAN, MCH-RZC test & $0.097$ & $0.680$ & $8.365$ & $0.029$ & $0.040$ & $0.024$ & $0.056$ & $0.501$ \\
        GAN, GOES-COT valid. & $0.140$ & $0.422$ & $8.652$ & $0.054$ & $0.059$ & $0.046$ & $0.073$ & $0.494$ \\
        GAN, GOES-COT test & $0.133$ & $0.456$ & $8.817$ & $0.061$ & $0.052$ & $0.044$ & $0.073$ & $0.506$  \\
        \hline
        GAN, MCH-RZC test & $0.097$ & $0.680$ & $\bm{8.365}$ & $\bm{0.029}$ & $\bm{0.040}$ & $\bm{0.024}$ & $\bm{0.056}$ & $\bm{0.501}$ \\
        Lanczos, MCH-RZC test & $0.092$ & $0.617$ & $18.700$ & --- & --- & --- & --- & --- \\
        RCNN, MCH-RZC test & $\bm{0.076}$ & $\bm{0.683}$ & $23.268$ & --- & --- & --- & --- & --- \\
        RainFARM, MCH-RZC test & $0.243$ & $0.134$ & $16.484$ & $0.131$ & $0.202$ & $0.318$ & $0.294$ & $0.516$ \\
        \hline
    \end{tabular}
    \caption{Image quality and variability metrics computed for the test and validation sets for the trained GAN.\newline
    Top: Metrics for the validation and testing sets of both the MCH-RZC and the GOES-COT datasets.\newline
    Bottom: Metrics for different methods using the MCH-RZC test set, bold numbers denoting the best method for each metric.}
    \label{table:metrics}
\end{table*}

The RMSE and MS-SSIM metrics improve rapidly in the first 15000 generator training sequences, converging quickly to a near-equilibrium. After this, there is little improvement in these scores. LSD keeps improving considerably longer especially for the MCH-RZC dataset, showing signs of improvement until approximately 70000 sequences. The CRPS metric, which utilizes all ensemble members, keeps improving longer than the single-image metrics, but with much more noise. After the switch to the SGD optimizer, the noise in the single-image metrics (but not the CRPS) is reduced, but the switch seems to have almost no effect on the metrics except for a slight degradation in the MS-SSIM metric for the GOES-COT dataset just after the switch.

Our subjective assessment of the generated image quality indicated that the quality keeps increasing for longer than the single-image metrics indicate, until at least 100000 sequences. We believe that the poor performance of the metrics is caused by them not capturing the desired qualities of the super-resolution reconstruction particularly well. The RMSE, in particular, is minimized at the mean of possible solutions, and therefore is of limited use in assessing the performance of GANs. The MS-SSIM is affected by similar issues because the objective of the GAN is to generate an ensemble of plausible solutions, and only a small fraction of those can be expected to be a close match to the original. For instance, when precipitation consists of small convective cells, the GAN might create cells of the correct size and intensity but in slightly wrong locations, leading to poor metrics in spite of perceptual similarity. The LSD, which compares the power spectra, does capture some of the structure but taking the power spectrum loses information about the location of the signals. The CRPS appears promising for evaluating conditional GANs as it detects improvement for much longer than the other metrics.

\subsection{Variability} \label{sect:variability}

\begin{figure}[!t]
\centering
\includegraphics[width=\linewidth]{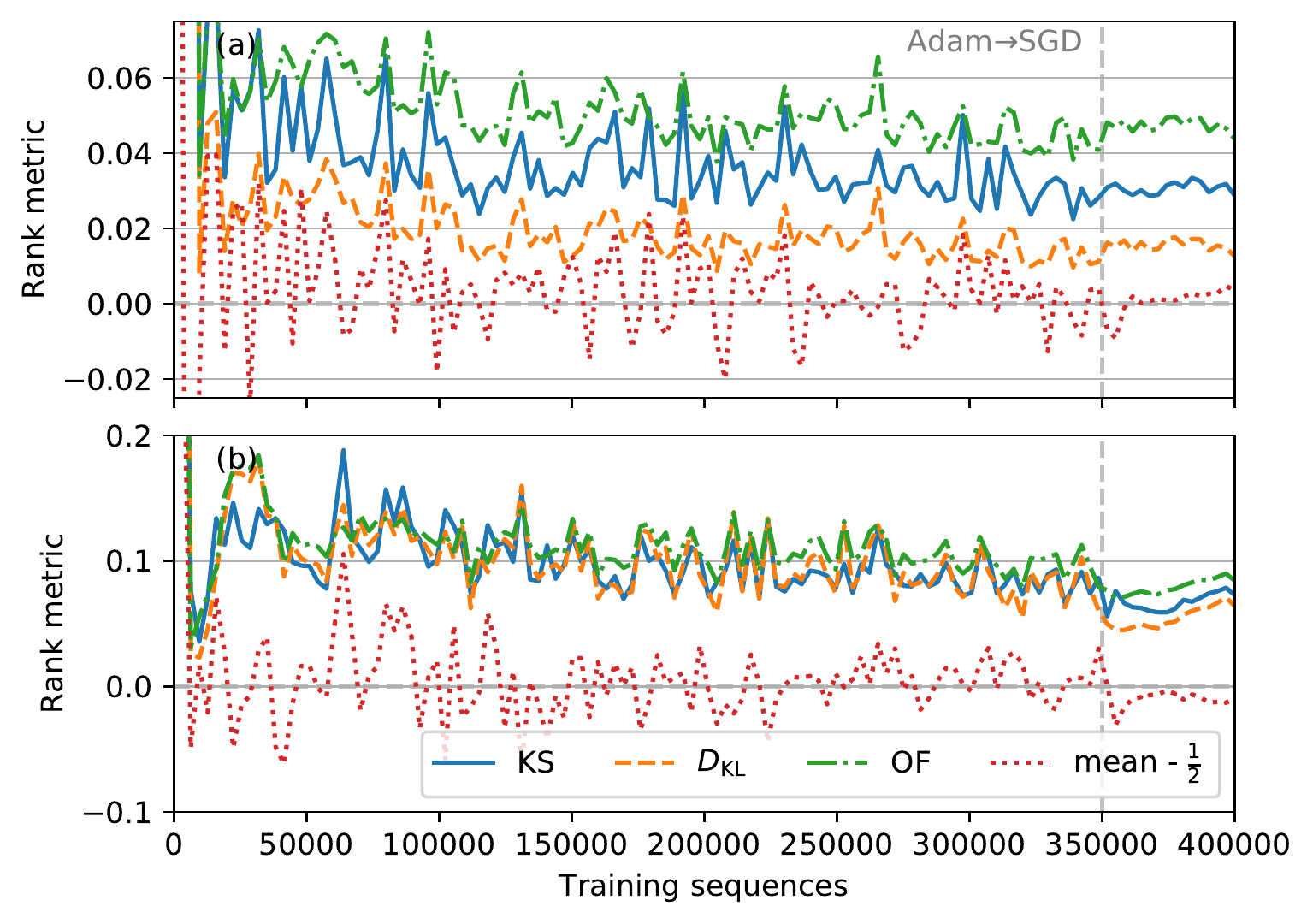}
\caption{Metrics of the rank distribution (as defined in Section~\ref{sect:validation}) of ground-truth images in the GAN-generated ensemble, shown as a function of training samples given to the generator. The blue solid line shows the Kolmogorov--Smirnov statistic, the orange dashed line shows the Kullback--Leibler divergence, the green dash-dotted line shows the outlier fraction, and the red dotted line shows the difference of the mean rank and $1/2$. Panel (a) shows the results for the MCH-RZC validation dataset and panel (b) for the GOES-COT validation dataset.}
\label{fig:rank-metrics-time}
\end{figure}

In Fig.~\ref{fig:rank-metrics-time}, we show the evolution of the variability metrics of the GAN over time during the training, evaluated using the validation dataset using $100$ ensemble members for each validation sample. We consider the KS statistic, $D_\mathrm{KL}$ and OF as defined in Section~\ref{sect:validation}, and also plot the bias of the mean rank from the optimal value of $1/2$. During the training using the Adam optimizer, the metrics improve rapidly at first, and slow improvement continues for much longer than with the single-image quality metrics discussed in the previous section. Improvement continues until at least 300000 sequences. After the switch to the SGD optimizer at approximately 350000 training sequences, the oscillation in the metrics is reduced.

\begin{figure}[!t]
\centering
\includegraphics[width=\linewidth]{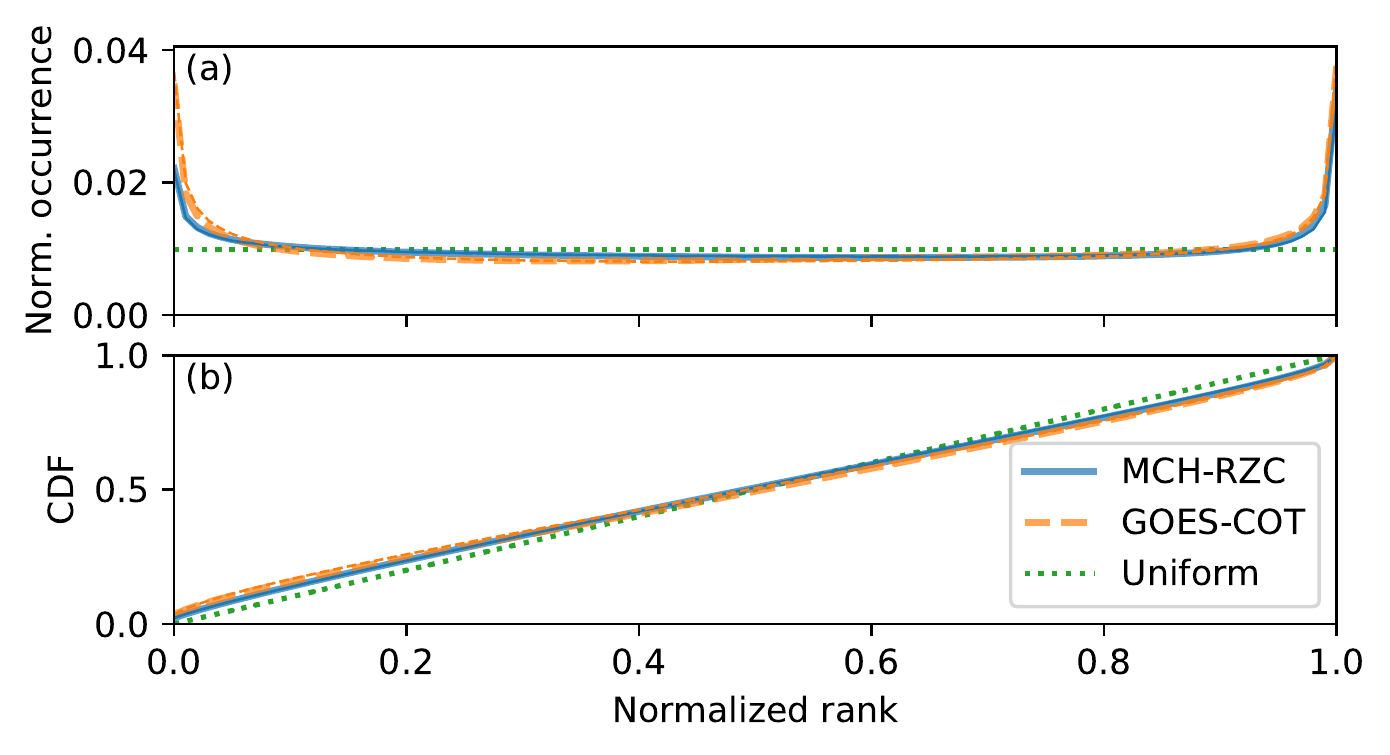}
\caption{(a) The occurrence of normalized ranks for the trained GAN (using the same generator weights as in Section~\ref{sect:examples}). The solid blue lines correspond to the MCH-RZC dataset and the dashed orange lines to the GOES-COT dataset. The thick, lighter-colored lines show the results for the test dataset, while the thin, darker lines show the results for the validation dataset. The green dotted line shows the uniform distribution for comparison. (b) As panel a, but showing the CDFs of the distributions.}
\label{fig:rank-distribution}
\end{figure}

The variability metrics for the fully trained GAN are shown in Table~\ref{table:metrics} alongside the quality metrics. The metrics near the end of training indicate that the rank distribution is close to uniform. At the time steps used in Section~\ref{sect:examples}, the KS statistic indicates that the CDF of the rank distribution differs from the CDF of the uniform distribution by at most $0.029$ for the MCH-RZC dataset, and at most $0.059$ for the GOES-COT dataset. This suggests that, at least in this respect, the GAN generates close to the appropriate amount of variability in its outputs, although there is clearly some difference in the distributions and therefore the actual KS test for equal distributions would fail.  The similarity to and differences from the uniform distribution can also be seen visually in Fig.~\ref{fig:rank-distribution} where we show the rank distribution graphically. The visualization shows that while there are considerably more samples in the outlier ranks ($r$ of either $0$ or $1$) than in the ranks near the middle of the distribution, these outliers represent only a minor fraction of all ranks (as also demonstrated by the OF in Table~\ref{table:metrics}). In a clear majority of cases, the real sample falls within the ensemble of predictions.

We also experimented with tuning the noise amplitude, which was noted by \cite{Leinonen2019CSMODISGAN} to increase the variability of the generated fields. We tried different multiplication factors for the noise, ranging from $0.5$ to $3.0$. We found that for inadequately trained generators, noise adjustment could significantly improve the variability metrics. On the other hand, for the models trained to near-convergence, the optimal adjustment factors were rather close to $1$, ranging between $0.9$ and $1.1$ depending on the dataset and the metric. Given that the difference is minor, and that there is no clear theoretical justification for this \textit{ad hoc} adjustment, we do not apply any adjustment to the noise amplitude in the final results.

\subsection{Comparison to alternative methods}  \label{sect:comparison}

\begin{figure}[!t]
\centering
\includegraphics[width=\linewidth]{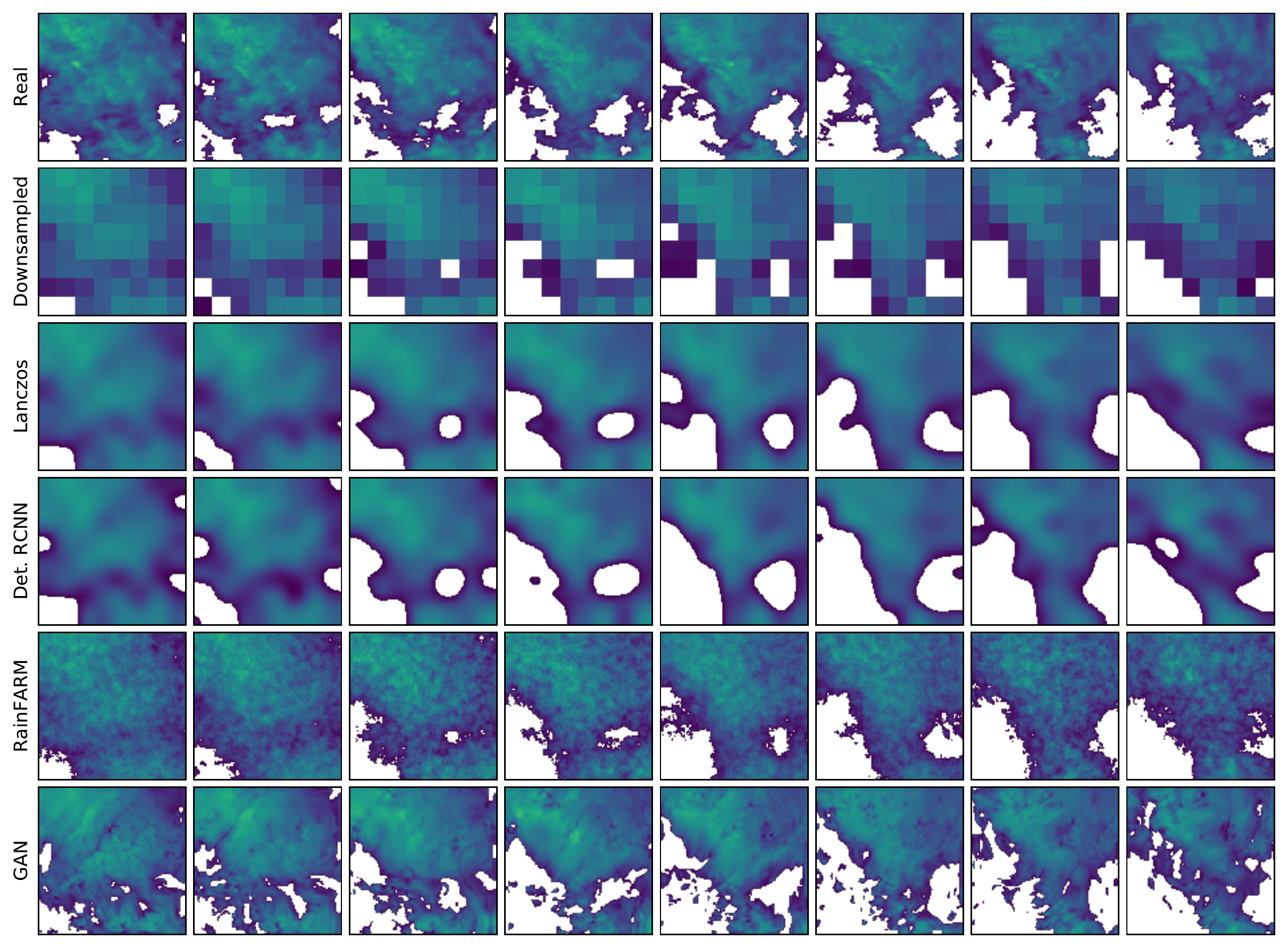}
\caption{Comparison of our GAN-based method to alternative methods. The first row shows the real high-resolution sequence and the second row shows the downsampled version. The subsequent rows show the different reconstruction methods: Lanczos interpolation (third row), RCNN trained to optimize RMSE (fourth row), the RainFARM algorithm (fifth row) and our GAN (sixth row).}
\label{fig:comparison}
\end{figure}

In Fig.~\ref{fig:comparison}, we show a comparison of our GAN-based method to alternative techniques: Lanczos interpolation, a recurrent CNN (RCNN) trained to optimize RMSE, and the Rainfall Filtered Autoregressive Model (RainFARM) algorithm of \cite{Rebora2006RainFARM}. These represent conceptually different approaches to the downscaling problem. Lanczos interpolation is a traditional, widely used image scaling method, and is used here as a baseline case. The RCNN trained with the RMSE loss is an example of a more straightforward deep-learning approach; in order to provide a fair comparison to the GAN, the RCNN uses the same architecture as our GAN generator, except with the noise input disabled. Finally, RainFARM is a downscaling method developed specifically for rainfall using more traditional statistical techniques based on Gaussian random fields generated using power-law spectral scaling. RainFARM, like the GAN, can be used stochastically to generate multiple realizations of the random field, while the RCNN and Lanczos methods are deterministic.

The examples illustrate that GAN produces more detail and a more visually accurate reconstruction of the original image than the alternative methods. The Lanczos interpolation and the RMSE-trained RCNN both produce a smooth output but with little detail at smaller scales. We also tried training the RCNN using the mean absolute error (MAE) loss, but the results (not shown) were very similar to RMSE. The RainFARM algorithm can produce more small-scale detail than the previous two methods, but it is limited to producing the same texture everywhere in the image and does not reproduce the structure of the high-resolution field as well as the GAN. Moreover, the example shown in Fig.~\ref{fig:comparison} is one where RainFARM performs relatively well. As the authors of \cite{Rebora2006RainFARM} note, RainFARM is quite sensitive to the choice of the scaling exponent, and we found that in some cases the textures produced were considerably less realistic than in this example as a result of a poorly estimated exponent. The GAN, on the other hand, works quite robustly and very rarely generates any implausible artifacts.

The performance metrics for the various methods are shown in the bottom half of Table~\ref{table:metrics}. These are consistent with what is shown in Fig.~\ref{fig:comparison}: The GAN, Lanczos and RCNN methods give similar results for the RMSE and MS-SSIM metrics, which further demonstrates that these are not particularly good metrics for evaluating GAN performance as they penalize solutions with higher variance. The RCNN achieves the best RMSE metric, which is unsurprising as it was specifically trained to optimize this metric, and it also gives the best MS-SSIM score. With the LSD, the GAN achieves the best score by far, while RainFARM, which produces a detailed texture, performs better than the Lanczos and the RCNN that produce unrealistically smooth outputs. In the ensemble metrics, the GAN clearly outperforms RainFARM, while these scores cannot be evaluated for the deterministic methods.

In terms of computational resources, evaluating the GAN generator (and, by extension, the RMSE-trained RCNN, which uses the same architecture) for one sequence of eight $128 \times 128$ pixel images took approximately $660\ \mathrm{ms}$ on a modern quad-core Intel i7 central processing unit (CPU) and $20\ \mathrm{ms}$ on an Nvidia P100 GPU. These times were obtained with a batch size of $16$; using a batch size of $1$ instead increased the evaluation time per sequence by approximately a factor of $2$ on both the CPU and the GPU (TensorFlow parallelization becomes less efficient with smaller batches). By comparison, the Lanczos interpolation took $11\ \mathrm{ms}$ seconds per sequence and the RainFARM algorithm, using a fairly unoptimized implementation, took $240\ \mathrm{ms}$ per sequence. The latter two methods were evaluated only on the CPU. This performance comparison demonstrates that the GAN method is relatively resource intensive but evaluating the GAN for modest amounts of input data is possible in a reasonable amount time also on a CPU, while a GPU is desirable for bulk processing large amounts of data.

\subsection{Generalization to larger images and longer sequences} \label{sect:generalization}

Since the GAN architecture is fully convolutional, we can apply the generator trained with relatively small (in our case $128 \times 128$ pixel) inputs to fields of different size without any modifications. The only restriction is that the pixels should correspond to the same physical size as the pixels of the training sequences, and that pixel dimensions of the input must be divisible by the resolution enhancement factor of $16$. Similarly, the recurrent structure allows us to apply the generator to longer or shorter sequences than the training sequences of length $8$ as long as the time interval between the frames of the sequence is the same as that used for training.

We demonstrated this capability by applying the generator (using the same version as in Section~\ref{sect:examples}) to the data from the June--August 2017 archive of full frames of the MCH-RZC data at $10\ \mathrm{min}$ time intervals. These data are from a different year than the training set and thus are completely independent. The frames in the data are $710$ pixels wide and $640$ pixels high; the width was cropped to $704$ pixels to satisfy the requirement that the dimensions be divisible by $16$. The generator was applied sequentially to each frame; the hidden state of the ConvGRU layer was propagated to the following frame at each step. For the first step, and wherever there is a longer than $10\ \mathrm{min}$ time gap between frames (which occasionally happens due to missing data), we used the initialization network to reinitialize the ConvGRU state, thus interrupting the time consistency in these situations.

\begin{figure*}[!t]
\centering
\includegraphics[width=\textwidth]{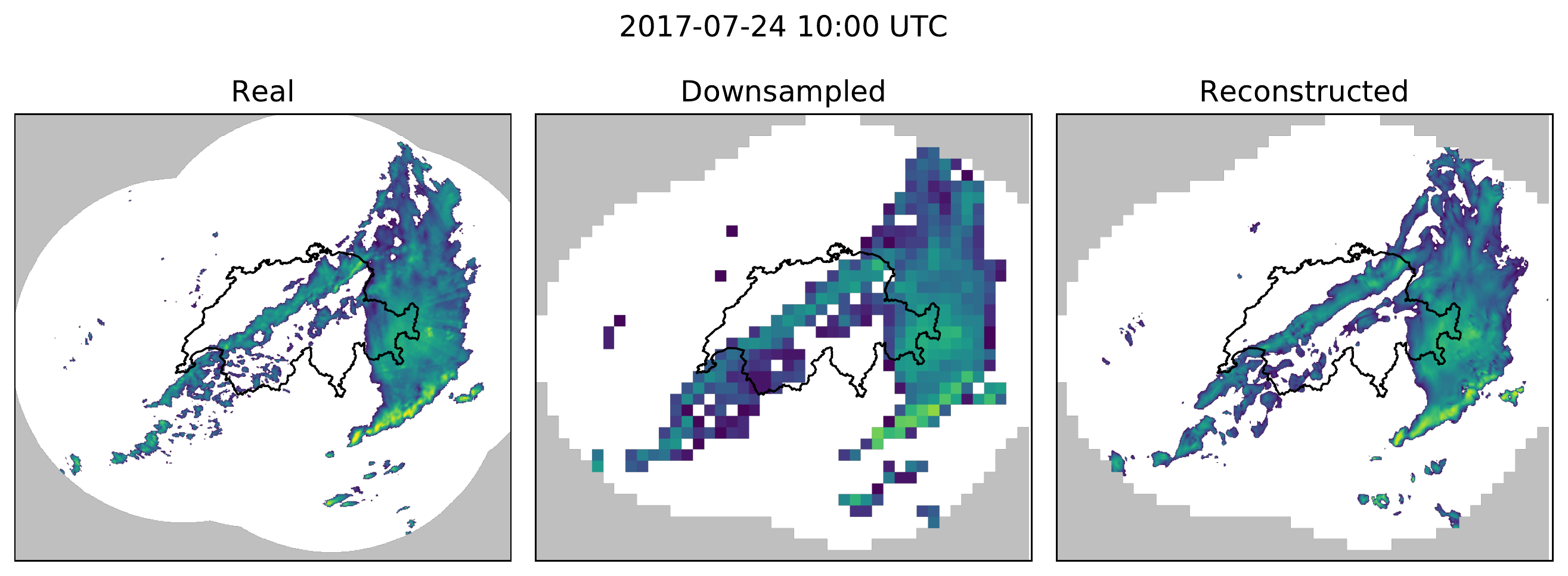}
\caption{An example of the results of the GAN applied to full frames of the June--August 2017 data from the MCH-RZC dataset, showing the situation of July 24 at 10:00 UTC. The gray areas mask the points that are unavailable due to lack of radar coverage. The borders of Switzerland are shown in the middle in order to provide spatial context. Left: the original frame. Middle: the downsampled version fed to the generator. Right: The high-resolution frame reconstructed by the GAN.}
\label{fig:mchrzc-fullframe}
\end{figure*}

We show one frame of the generated sequence in Fig.~\ref{fig:mchrzc-fullframe}. This example shows a situation with different modes of precipitation in different regions. It demonstrates that the GAN can create realistic reconstructions even for much larger images than those from the training set. The time evolution of the generated fields can obviously not be properly illustrated with a single image, so we provide an animation that shows the June--August 2017 sequence as a video accompanying this article online. 

While generating these long time series, we found that some versions of the generator could produce artifacts when left running for a long time. For the purposes of generating Fig.~\ref{fig:mchrzc-fullframe} and the corresponding video, we were able to suppress these artifacts sufficiently by adjusting the generator architecture and choosing a version of the generator that was less prone to them. However, for those cases where the artifacts cannot be avoided, we found a simple stabilization method, which we describe in the Appendix.

\section{Summary and Conclusions} \label{sect:conclusions}

Deep learning has enabled significant advances in image and video super-resolution, with GANs being among the most prominent methods. Resolution enhancement also has many applications in the processing of observational and model data in the weather and climate sciences. However, in weather and climate applications, uncertainty quantification is essential. The present work addresses this need with a conditional super-resolution GAN that operates on sequences of two-dimensional images and creates an ensemble of predictions for each input. The spread between the ensemble members represents the uncertainty of the super-resolution reconstruction.

Rather than processing each image in a sequence independently, our generator architecture uses a recurrent layer to update the state of the high-resolution reconstruction in a manner that is consistent with both the previous state and the newly received data. The recurrent layer can thus be understood as performing a Bayesian update on the ensemble member, resembling an ensemble Kalman filter. Besides being recurrent, the generator is fully convolutional, meaning that it can operate on variable-sized inputs and produce consistent time evolution for arbitrarily long sequences.

The representativeness of the ensemble was quantitatively evaluated using ensemble statistics. We found that rank metrics take longer to converge than image quality metrics such as MS-SSIM and RMSE, and therefore the rank metrics can be used to monitor the progress of the training even after image quality metrics saturate. The CPRS quality metric, which uses the entire ensemble, also appears to provide a better estimate of image quality than the single-image metrics. The ensemble metrics therefore seem promising for evaluating the quality and variability produced by conditional GANs in general and may be useful in applications beyond the geoscience domain.

The evaluation of the GAN indicates that it produces realistic high-resolution fields with appropriate amounts of variability. Moreover, the GAN was trained separately for two distinct applications, proving that it can generalize to different types of input data. We expect that it can be applied to other similar applications as well. The GAN generator also generalizes well to larger input images and longer sequences than those in the training set, reducing the computational cost of training as the GAN can be trained with relatively short sequences of small images and then evaluated with sequences of different length and image size.

Besides increasing the range of applications, potential future improvements include:
\begin{itemize}
\item Generalization to different scaling factors or possibly producing high-resolution images for multiple scaling factors at once (the current version is specific to the factor of $16$).
\item Resolution enhancement in the temporal as well as the spatial dimension to allow time interpolation.
\item The inclusion of auxiliary variables to help the generator produce the right kind of fields; for instance, orography affects precipitation formation  and could be included as an additional variable, as was previously done in a deep-learning context by \cite{Franch2020Nowcasting}.
\item Further development of the rank-based methods for evaluating conditional GANs. In particular, the ensemble metrics in this paper were evaluated pixelwise, but it may be possible to develop a more feature-based method similar to the FID.
\end{itemize}

\appendix[Optional stabilization for long time series]

We found that some versions of the generator were prone to generating artifacts when left running recurrently for many time steps. In these cases, the generator was stable over the $8$ frames used in the training, but this was apparently not always sufficient to guarantee stability over longer periods of time. While we were able to avoid this in our reported experiments, as described in Section~\ref{sect:generalization}, we found a relatively simple technique to suppress the artifacts when they appear. We report it here as it may be useful for further experiments with such recurrent GANs.

As the initialization network did not produce any artifacts, we were able to use the following procedure to stabilize the evaluation of the generator: On each time step $k$, after evaluating the update network, the ConvGRU state $h_k$ is adjusted as follows:
\begin{equation}
h_k \coloneqq h_\mathrm{null} + (1-\lambda_r)(h_k-h_\mathrm{null})
\end{equation}
where $h_\mathrm{null}$ is the ConvGRU state produced by the initialization network for an all-zeros input, and $\lambda_r$ is a relaxation constant (we experimented with $0.01 \leq \lambda_r \leq 0.2$ for the MCH-RZC dataset). This process nudges the ConvGRU state towards the null state. This effectively suppresses artifacts while still allowing the update network to operate on the state from the previous step. This procedure seems to reduce (but not completely eliminate) the variability present in the generated images. Therefore, while it serves to stabilize the evaluation over long periods of time, it should only be used when the artifacts cannot be removed using improvements to the generator network.


%

\appendices


\section*{Acknowledgment}
The authors acknowledge the financial support from the Swiss National Science Foundation (grant \#200020\_175700). The GPU computing resources used in this study were provided through a grant from the Swiss National Supercomputing Centre (CSCS) under project ID sm35. The authors would like to thank MeteoSwiss for providing the MCH-RZC data and Dr. D. Wolfensberger for assisting us with using it.

\ifCLASSOPTIONcaptionsoff
  \newpage
\fi



\bibliographystyle{IEEEtranDOI}
\bibliography{journalabrv,srrefs}
%



%

\begin{IEEEbiography}[{\includegraphics[width=1in,height=1.25in,clip,keepaspectratio]{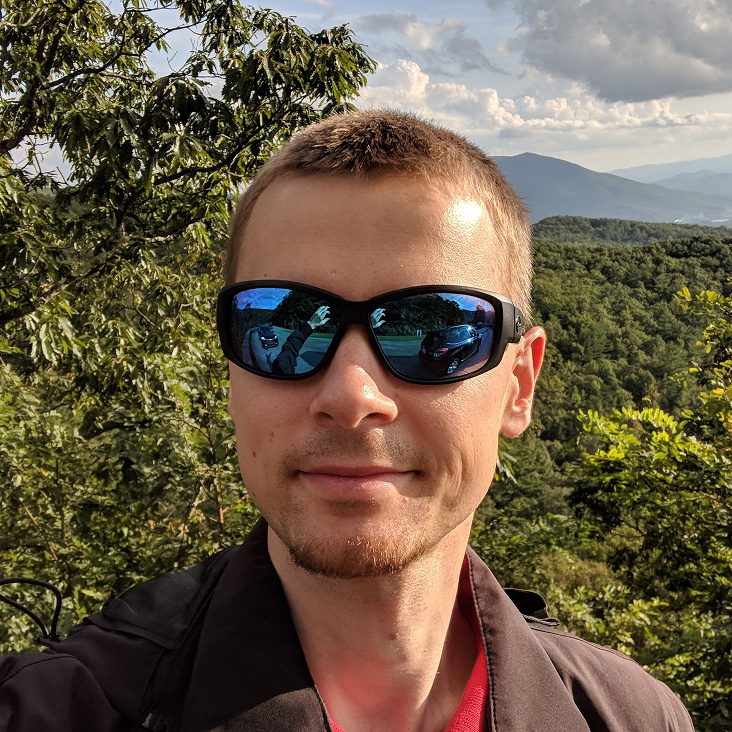}}]{Jussi Leinonen}
received the M.S. degree from the Helsinki University of Technology in Espoo, Finland, in 2007 and the Dr. Tech. Sci. degree from the Aalto University in Espoo, Finland, in 2013. He performed his doctoral research at the Finnish Meteorological Institute in Helsinki, Finland, and afterwards was a Postdoctoral Scholar and later a Data Scientist at the Jet Propulsion Laboratory, California Institute of Technology, in Pasadena, California, USA, between 2014 and 2019. Since April 2019, he has been a Scientist at the Environmental Remote Sensing Group, École polytechnique fédérale de Lausanne, in Lausanne, Switzerland.
His research interests include machine learning in the atmospheric sciences, precipitation radars, electromagnetic scattering, cloud and precipitation microphysics and probabilistic data analysis.
\end{IEEEbiography}

\begin{IEEEbiography}[{\includegraphics[width=1in,height=1.25in,clip,keepaspectratio]{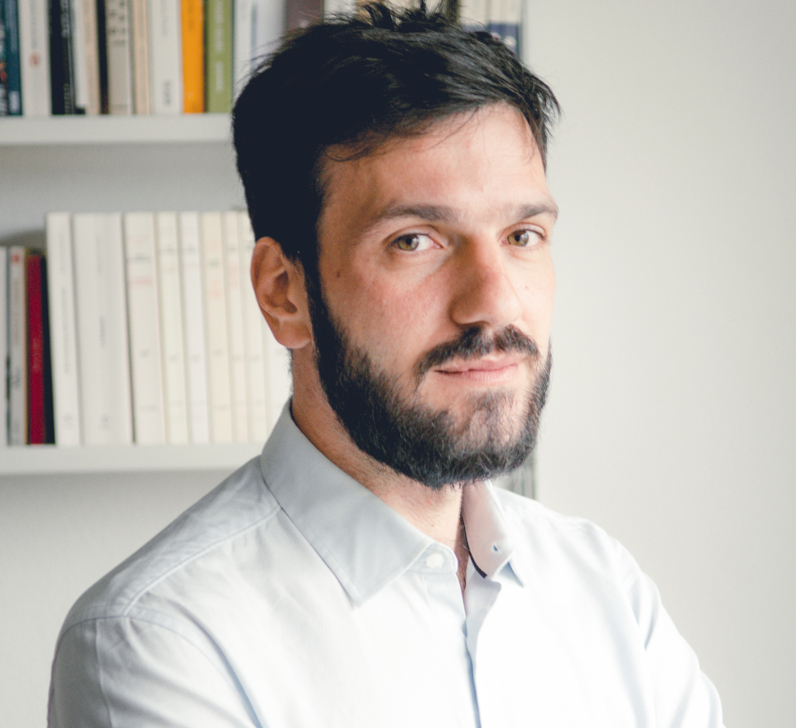}}]{Daniele Nerini} is currently employed as a research associate in the Forecast Development Division at the Swiss Federal Office for Meteorology and Climatology MeteoSwiss in Locarno-Monti, Switzerland. He obtained the M.S. degree from Imperial College London, England, in 2013 and the PhD from the ETH Zurich, Switzerland, in 2019. He performed his doctoral research in the Radar, Satellite, and Nowcasting Division at MeteoSwiss. His current research focuses on radar hydrology, precipitation nowcasting, and the verification and post-processing of numerical weather predictions. 

\end{IEEEbiography}

\begin{IEEEbiography}[{\includegraphics[width=1in,height=1.25in,clip,keepaspectratio]{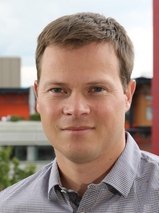}}]{Alexis Berne}
Alexis Berne received the Ph.D. degree from Université Joseph Fourier, Grenoble, France, in 2002. From 2003 to 2006, he was a Marie Curie Fellow with Wageningen University, Wageningen, The Netherlands. Since 2006, he has been leading the Environmental Remote Sensing Laboratory at the École Polytechnique Fédérale de Lausanne, Lausanne, Switzerland. His current research interests include radar meteorology in mountainous and polar regions.
\end{IEEEbiography}





\end{document}